\documentclass{article}
\usepackage{epsfig}

\date{\today}

\begin{document}

\begin{center}
{\Large\bf Magnetic charge, angular momentum and negative cosmological constant}
\vspace{1.cm}\\
J.J.\ van der Bij and Eugen Radu\footnote{\textbf{corresponding author}:
\\Albert-Ludwigs-Universit\"at Freiburg, 
\\Fakult\"at f\"ur Mathematik und Physik,
Physikalisches Institut,
\\Hermann-Herder-Stra\ss e 3, D-79104 Freiburg, Germany
\\ email: radu@newton.physik.uni-freiburg.de
\\ Telephone: +49-761/203-7630 
\\Fax: +49-761/203-5967}\vspace{0.4cm}\\
\it Albert-Ludwigs-Universit\"at \\
\it Fakult\"at f\"ur Mathematik und Physik,
\\
Physikalisches Institut,
\it Freiburg Germany\vspace{0.6cm}\\
\end{center}
\begin{abstract}
We argue that there are no  axially symmetric 
rotating monopole solutions for a Yang-Mills-Higgs theory 
in flat spacetime background.
We construct axially symmetric  Yang-Mills-Higgs solutions 
in the presence of a negative 
cosmological constant, carrying magnetic charge $n$ and a 
nonvanishing electric charge.
However, these solution are also nonrotating. 
\end{abstract}

\section{Introduction}
Even if nobody will ever see a magnetic monopole, there is surely much
to be gained by studying the theory of monopoles.
In the seminal paper of Julia and Zee \cite{Julia-Zee}
the existence of a profound  connection 
between the angular momentum and the electric charge 
in a Yang-Mills-Higgs  (YMH) theory has been suggested.
A well know result in classical electrodynamics implies 
that in the presence of a magnetic monopole an electrically charged particle
acquires an additional angular momentum oriented toward the position of the monopole 
and with a magnitude equal to the product of the electric and magnetic charges \cite{Jackson}.
In the nonabelian case the electric charge distribution and the magnetic charge distribution 
are not separated. To compute the angular momentum for 
this type of  configuration
we have to solve extremely complicated partial differential 
equations.
 
Surprisingly, a few years ago it has been shown that Julia-Zee dyons do not
admit slowly rotating excitations \cite{HSV}. However the absence
or presence of rotating  perturbative solutions, though
indicative, is not in general conclusive to establish the absence or
presence of exact solutions.  For instance in boson star models slowly rotating
perturbative solutions are absent, but solutions with a quantized angular
momentum exist \cite{boson} (see also the discussion in \cite{Volkov:2002aj}). 
It is therefore important to
identify non-perturbative criteria to settle such questions. 

In a recent paper a general 
relation for the total angular momentum in an Einstein-Yang-Mills-Higgs theory
as a surface integral in terms of Yang-Mills (YM) fields 
has been derived \cite{VanderBij:2001nm}.
Within an ansatz which satisfies some extra symmetries it has been found that
there are no asymptotically flat rotating dyon solutions, 
while a monopole-antimonopole configuration with zero net magnetic charge possesses
an angular momentum proportional to the electric charge.
However Ref.\cite{VanderBij:2001nm} leaves unsolved the 
question of possible existence of finite energy solutions 
with a nonvanishing total angular momentum and net magnetic charge
for a more general ansatz.

In this paper we argue that, for the most general axially symmetric
YMH monopole configuration, the total angular momentum is zero.
We adress also the same question for a different asymptotic structure 
of spacetime and
present arguments for the existence of axially symmetric dyon solutions
in an anti-de Sitter (AdS) spacetime.
Qualitatively, the behavior of AdS solutions is very 
similar to that corresponding
to Minkowski spacetime configurations.
However, we argue that these solutions present also a vanishing total angular momentum.

\section{General ansatz and total angular momentum}
To simplify the general picture, in this paper we ignore  the effects of gravity.
We follow also the notations and conventions 
used by Hartmann, Kleihaus and Kunz in \cite{Hartmann:2001ic}.
The action for a  non-Abelian $SU(2)$ gauge 
field coupled to 
a triplet Higgs field with the usual potential 
$V(\Phi)=\frac{\lambda}{8} Tr(\Phi^2 - \eta^2)^2$  is
\begin{equation} 
\label{lag0}
S=\int d^{4}x\sqrt{-g}[
Tr\{\frac{1}{2}F_{\mu \nu}F^{\mu \nu}\}+Tr\{\frac{1}{2}D_{\mu}\Phi D^{\mu}\Phi\}+
V(\Phi)],
\end{equation}
where the field strength tensor is
\begin{eqnarray}
\nonumber
F_{\mu \nu}&=&\partial_{\mu}A_{\nu}-\partial_{\nu}A_{\mu}+i[A_{\mu},A_{\nu} ],
\end{eqnarray}
and the covariant derivative
\begin{eqnarray}
D_{\mu}&=&\partial_{\mu}+i[A_{\mu},\ ].
\nonumber
\end{eqnarray}
Varying the action (\ref{lag0}) with respect to
$A_{\mu}$ and $\Phi$ 
we have the field equations 
\begin{eqnarray}
\label{YMeqs}
\frac{1}{\sqrt{-g}}D_{\mu}(\sqrt{-g} F^{\mu \nu}) &=&\frac{1}{4} i[\Phi,D^{\nu}\Phi],
\\
\label{Heqs}
\frac{1}{\sqrt{-g}}D_{\mu}(\sqrt{-g}D^{\mu} \Phi) +\lambda(\Phi^2-\eta^2)\Phi&=&0.
\end{eqnarray}
Varying the Lagrangian with respect
to the metric $g_{\mu \nu}$ yields the energy-momentum tensor
\begin{eqnarray}
\label{tensor}
T_{\mu\nu} = 2Tr\{F_{\mu \alpha} F_{\nu \beta} g^{\alpha \beta} 
-\frac{1}{4}g_{\mu \nu}F_{\alpha \beta}F^{\alpha \beta} \}
+Tr\{\frac{1}{2}D_{\mu}\Phi D_{\nu}\Phi 
-\frac{1}{4}g_{\mu \nu}(D_{\alpha}\Phi D^{\alpha}\Phi\}+
V(\Phi)g_{\mu\nu}.
\end{eqnarray}
The energy density of a solution of the YMH equations 
is given by the $tt-$component of the 
energy-momentum tensor. Integration over all space yields the total energy
\begin{eqnarray}
\label{total-energy}
E &=& -\int T_{t}^{t}\sqrt{-g} d^{3}x
=\int_{0}^{\infty}dr \int_{0}^{\pi}d\theta \int_{0}^{2\pi}d\varphi \sqrt{-g}
Tr \Big\{
F_{r\theta}F^{r\theta}+F_{r\varphi}F^{r\varphi}+F_{\theta \varphi}F^{\theta \varphi}
\\
\nonumber
&-&F_{rt}F^{rt}-F_{\theta t}F^{\theta t}-F_{\varphi t}F^{\varphi t}
+\frac{1}{4}(D_r\Phi D^r\Phi+D_{\theta}\Phi D^{\theta}\Phi+D_{\varphi}\Phi D^{\varphi}\Phi
-D_t\Phi D^t\Phi)+\frac{\lambda}{8} (\Phi^2 - \eta^2)^2
\Big\}
,
\end{eqnarray}
where $r,~\theta,~\varphi$ are the usual spherical coordinates.
For the energy integral to converge, each of the terms in (\ref{total-energy}) 
must vanish at large $r$.
The finite energy condition imposed on this expression will give us the acceptable 
asymptotic behavior 
of the gauge functions and Higgs field as $r \to \infty$.
The additional requirements to have a finite, locally integrable energy density
 impose boundary conditions 
at the origin and on the $z-$axis.

The total angular momentum is
\begin{eqnarray}
\label{J}
J&=&\int T_{\varphi}^{t}\sqrt{-g} d^{3}x
= \int 2Tr\{F_{r \varphi} F^{r t}
+F_{\theta \varphi} F^{\theta t}+D_{\varphi}\Phi D^{t}\Phi\} \sqrt{-g} d^{3}x.
\end{eqnarray}
We mention also the expression for the electric and magnetic charges
derived by using the 't Hooft field strength tensor
\begin{eqnarray}
\label{e-charge}
Q_e&=&\frac{1}{4\pi}\oint_{\infty}dS_{\mu}Tr\{ \hat{\Phi}F_{\mu t} \},
\\
\label{m-charge}
Q_m&=&\frac{1}{4\pi}\oint_{\infty}dS_{\mu}\frac{1}{2}\epsilon_{\mu \nu \alpha}
Tr\{\hat{\Phi}F_{\nu \alpha}\},
\end{eqnarray}
where the integration is over a surface at spatial infinity and $\hat{\Phi}=\Phi/|\Phi|$.

We are interested in static, axially symmetric finite energy solutions of the equations 
(\ref{YMeqs})-(\ref{Heqs}).
As proven in \cite{VanderBij:2001nm}, for static axially symmetric configurations,
the volume integral (\ref{J}) can be converted 
into a surface integral in terms of Yang-Mills potentials. 
The symmetry of the gauge field under a spacetime symmetry means that
the action of an isometry can be compensated by a suitable gauge transformation
\cite{Heusler:1996ft,Forgacs:1980zs}. 
For the time translational symmetry, we choose a natural gauge such that $\partial A/\partial t$=0.
However, a rotation around the $z-$axis 
can be compensated by a gauge rotation
\begin{eqnarray} \label{Psi}
{\mathcal{L}}_\varphi A=D\Psi,
\end{eqnarray}
and therefore
\begin{eqnarray} 
\label{relations}
F_{\mu \varphi} =& D_{\mu}W,
\\
D_{\varphi}\Phi=&i[W,\Phi] \nonumber,
\end{eqnarray}
where $W=A_{\varphi}-\Psi$.

Using the potential $W$ we find
\begin{eqnarray}
T_{\varphi}^{t}&=&2Tr\Big\{ (D_{r}W)F^{rt}
+(D_{\theta}W)F^{\theta t}
+\frac{i}{4}[W,\Phi] D^{t} \Phi\Big\}
\nonumber\\
&=&2Tr\Big\{\frac{1}{\sqrt{-g}}D_{r}(WF^{rt}\sqrt{-g})
+\frac{1}{\sqrt{-g}}D_{\theta}(WF^{\theta t}\sqrt{-g})
\nonumber\\
&^{}&-W\Big(
\frac{1}{\sqrt{-g}}D_{r}(\sqrt{-g}F^{rt})
+\frac{1}{\sqrt{-g}}D_{\theta}(\sqrt{-g}F^{\theta t})\Big)
+\frac{i}{4}[W,\Phi] D^{t} \Phi\Big\}.
\end{eqnarray}
As a consequence of the YM equations (\ref{YMeqs}) 
and making use of the fact that the trace 
of a commutator vanishes we obtain 
\begin{eqnarray}
\label{T34}
T_{\varphi}^{t}=2Tr\Big\{\frac{1}{\sqrt{-g}}\partial_{\mu}(WF^{\mu t}\sqrt{-g})\Big\}.
\end{eqnarray}
Thus, ignoring possible singularities the expression of the total angular momentum is
\begin{eqnarray}
\label{totalJ}
J &=&\oint_{\infty}2Tr\{WF^{\mu t} \} dS_{\mu}
\nonumber\\
&=&2 \pi \lim_{r \rightarrow \infty}  \int_0^{\pi} d \theta \sin \theta^{~}
 r^2[W^{(r)}F^{rt(r)}+W^{(\theta)}F^{rt(\theta)}+W^{(\varphi)}F^{rt(\varphi)}]. 
\end{eqnarray} 
Note that this relation has been derived without fixing the gauge.
It greatly simplifies the evaluation of the angular momentum of a 
regular YMH configuration, since we need 
only the asymptotics of some gauge functions. 
In what follows, we use (\ref{totalJ}) to prove that the total angular momentum 
of a finite energy dyon solution in flat space is zero.

The method of proof is, in principle, quite straightforward.
From the assumptions of finite energy density and finite total energy, it follows
that every term in the energy integral (\ref{total-energy}) must be finite.
The most crucial requirement is that $D_{\mu}\Phi$
vanish sufficiently rapidly as $r \to \infty$. 
When evaluating (\ref{totalJ}) we find a vanishing $J$ for this behavior.

The construction of an axially symmetric YMH ansatz has been discussed by many authors 
starting with the pioneering papers by Manton \cite{Manton:1977ht} 
and Rebbi and Rossi \cite{Rebbi:1980yi}. 
The most general axially symmetric Yang-Mills ansatz contains 12 functions: 9 magnetic
and 3 electric potentials and can be easily obtained in cylindrical coordinates 
$x^{\mu}=(\rho,\varphi,z,t$) (with $\rho=r\sin \theta,~z=r\cos\theta)$
\begin{eqnarray} 
\label{A-gen-cil}
A_{\mu}=\frac{1}{2}A_{\mu}^{\rho}(\rho,z)\tau_{\rho}^n
        +\frac{1}{2}A_{\mu}^{\varphi}(\rho,z)\tau_{\varphi}^n
        +\frac{1}{2}A_{\mu}^{z}(\rho,z)\tau_{z},
\end{eqnarray}
where the only $\varphi$-dependent terms are the $SU(2)$ matrices
(composed of the standard $(\tau_x,~\tau_y,~\tau_z)$ Pauli matrices)
\begin{eqnarray} 
\label{u-cil}
\tau_{\rho}^n&=&~~\cos n\varphi~\tau_x+\sin n\varphi~\tau_y,
\nonumber\\
\tau_{\varphi}^n&=&-\sin n\varphi~\tau_x+\cos n\varphi~\tau_y.
\end{eqnarray}
We can reduce this general ansatz by imposing some extra symmetries on the fields.
To our knowledge there are no analytical or numerical results
for the most general ansatz.

Transforming to spherical coordinates, it is convenient to introduce,
without any loss of generality, a new $SU(2)$ basis 
$(\tau_{r}^n,\tau_{\theta}^n,\tau_{\varphi}^n)$,
with
\begin{eqnarray} 
\label{u-sph}
\tau_{r}^n&=&\sin \theta~\tau_{\rho}^n+\cos \theta~\tau_z,
\nonumber\\
\tau_{\theta}^n&=&\cos \theta~\tau_{\rho}^n-\sin \theta~\tau_z.
\end{eqnarray}
Thus, similar to (\ref{A-gen-cil}), we find the general expression
\begin{eqnarray} 
\label{A-gen-sph}
A_{\mu}=\frac{1}{2}A_{\mu}^{r}(r,\theta)\tau_{r}^n
        +\frac{1}{2}A_{\mu}^{\theta}(r,\theta)\tau_{\theta}^n
        +\frac{1}{2}A_{\mu}^{\varphi}(r,\theta)\tau_{\varphi}^n,
\end{eqnarray}
where
$A_\mu^a dx^\mu=A_r^a dr + A_\theta^a d\theta + A_\varphi^a d\varphi+A_t^a dt$.
For the Higgs field we have a similar relation $\Phi=\Phi^a(r,\theta) \tau_a^n$.
The gauge invariant quantities expressed in terms of these functions 
will be independent of the azimutal
angle $\varphi$ and hence axially symmetric.
The winding number $n$ corresponds to the topological 
charge of the solutions and is a constant of motion.

For this parametrization 
\begin{equation}
\label{psi}
\Psi=n\frac{\tau_z}{2}=n \cos \theta \frac{\tau_{r}^n}{2} - n \sin \theta \frac{\tau_{\theta}^n}{2}.
\end{equation}
The Higgs field must be nonvanishing at spatial infinity in order that the potential
energy be zero there. Thus, the spontaneous symmetry-breaking mechanism requires
\begin{equation}
\label{Hinf} 
\lim_{r \rightarrow \infty}\Phi^a\Phi^a=\eta^2.
\end{equation}
As proven in \cite{Houston:aj}, any regular axially symmetric magnetic charge distribution
can be located only at isolated points situated 
on the axis of symmetry, with equal and opposite values of the charge at alternate points.
In particular, if only one sign of the charge is allowed,
all the charge must be concentrated at a single point.

For a multimonopole solution containing only magnetic charges with the same sign, 
the suitable boundary conditions for the Higgs field at 
infinity are $\Phi^r=\eta,~
\Phi^{\theta}=\Phi^{\varphi}=0$. Note that a more general asymptotic 
behavior of the Higgs field is allowed by (\ref{Hinf}), but this would correspond to composed 
configurations containing both monopoles and antimonopoles.

The requirement of finite total energy implies the decay of the $F_{rt}^a$ 
as $r \to \infty$ is faster than $1/r^{1.5}$.
Also, the contribution $(D_{\varphi}\Phi)^a (D_{\varphi}\Phi)^a g^{\varphi \varphi}$ term
to the energy is finite if $(D_{\varphi}\Phi)^a$
approaches zero to large $r$ sufficiently rapid.
This implies the fall-off conditions $(W_{\theta}, W_{\varphi}) \sim 1/r^{0.5+\epsilon}$ for large $r$ 
(with $\epsilon>0$).
Therefore the last two terms in (\ref{totalJ}) give null contribution to the total angular
momentum.
The contribution of the $A_{\varphi}^r$ to the total angular 
momentum is also zero, since it vanishes asymptotically
for a regular configuration.  

By using these results we find 
\begin{equation} 
\label{res-dyon} 
 \lim_{r \rightarrow \infty}Tr (r^2 WF_{r t})=-\frac {n  Q_e}{2}  \cos \theta 
\end{equation} 
where $Q_e$ is the electric charge of the configuration given by (\ref{e-charge}). 
Therefore the total angular momentum of the solution is clearly zero. 
Thus, an axially symmetric regular dyon has a vanishing total angular momentum.

\section{Axially symmetric dyons in AdS background}
This nonexistence result can be partially attributed to the asymptotic structure 
of spacetime.
One may ask whether it remains valid in the presence of 
a cosmological constant $\Lambda$.

However, up to date, all known axially symmetric dyon regular solutions 
have been obtained in a fixed Minkowski background.
In what follows we consider dyon solutions 
in an AdS spacetime. 
Further motivation for this attempt
comes from the surprising behavior found
for a $SU(2)-$gauge field in the presence of a negative cosmological constant.
The existence of stable particle-like and black hole solutions 
in an Einstein-Yang-Mills theory with $\Lambda<0$ is the most exciting result
\cite{Winstanley:1999sn, Bjoraker:2000qd}.

The global existence of a solution of the Cauchy problem for the YMH equations in
AdS spacetime is considered in \cite{Choquet-Bruhat:1989bn}.
Some analytical properties of spherically symmetric monopole and 
dyon solutions for YMH theory with a negative cosmological constant
have been discussed in \cite{Lugo:1999fm}. Unlike the flat spacetime case, 
in an AdS spacetime there are no analytic solutions that might be used as a guiding line.
Moreover, when a cosmological constant is included (no matter how small this constant is)
no solution close to the BPS configuration can be found.
As  proven in \cite{Lugo:1999fm}, it is the change in the asymptotic 
behavior of the Higgs field that prevents such a solution.
Numerical monopole solutions have been exhibited in \cite{Lugo:1999ai}, 
where the effects of gravity are also included.
A distinctive feature of AdS solutions concerns the asymptotic behavior of the fields. 
When $\Lambda<0$, the Higgs field approaches its vacuum expectation value faster than
in the flat space case. The radius of the monopole core decreases, 
as the magnetic field concentrates near the origin.

In what follows, we present numerical arguments for the existence of spherically- and
axially symmetric dyon solutions in an AdS spacetime with line-element 
\begin{equation} 
\label{AdS}
ds^2= \frac{d r^2}
{1-\frac{\Lambda}{3}r^2}+ r^2 (d \theta^2
           + r^2 \sin ^2 \theta d\varphi^2)- (1-\frac{\Lambda}{3}r^2) dt^2.
\end{equation} 
Searching for general axially symmetric solutions within the most general ansatz
is a difficult task.
In our approach we consider a reduced ansatz 
used also to obtain flat space - axially symmetric dyons  \cite{Hartmann:2000ja}  
\begin{eqnarray}
\label{ansatz-AdS}
\nonumber
A_{r}^r&=&A_{r}^{\theta}~=~A_{\theta}^r~=~A_{\theta}^{\theta}
~=~A_{\varphi}^{\varphi}=\Phi^{\varphi}=0,
\\
A_r^r&=&\frac{H_1}{r},~A_{\theta}^r~=~1-H_2,~A_{\varphi}^r~=~-n \sin \theta H_3,~
A_{\varphi}^{\theta}~=~-n \sin \theta (1-H_4),
\\
\nonumber
A_t^r&=&\eta H_5,~A_t^{\theta}~=~\eta H_6,~\Phi^r~=~\eta \Phi_1,~\Phi^{\theta}~=~\eta \Phi_2.
\end{eqnarray}
This ansatz satisfies some discrete symmetries \cite{Brihaye:1994ib}
and was used by various authors when discussing numerically solutions \cite{Hartmann:2001ic}.
The boundary conditions at infinity consistent with the requirements
of regularity, finite energy and symmetry, are
\begin{equation} 
\label{cond-inf}
H_{i}=0, \ i=1,2,3,4,6; \ \ \ \
H_{5}=\alpha;\ \ \ \
\Phi_{1}=1;\ \ \ \ \Phi_{2}=0
\end{equation}
at infinity, and
\begin{equation} 
\label{cond-orig}
H_{i}=0, \ i=1,3,5,6; \ \ \
H_{2}=H_{4}=1, \ \ \
\Phi_{1}=\Phi_{2}=0,
\end{equation} 
at the origin.
Given the parity reflection symmetry, we need to consider solutions 
only in the region $0 \leq \theta \leq \pi/2$; 
on the $z$- and $\rho$-axis the functions
$H_1, H_3, H_6, \Phi_2$ and the derivatives 
$\partial_\theta H_2,\partial_\theta H_4,
\partial_\theta H_5$ and $\partial_\theta \Phi_1$ are to vanish.
To fix the residual abelian gauge invariance we choose the usual gauge condition 
$ r \partial_r H_1 - \partial_\theta H_2 = 0$ \cite{Hartmann:2001ic,Hartmann:2000ja}.

Within this ansatz, we have found dyon solutions for any value of the cosmological constant.
We change to dimensionless coordinates and Higgs field by rescaling
$r \to r /\eta $, $\Lambda \to \Lambda \eta^2$ 
and $\Phi \to \eta \Phi$, respectively.
The numerical calculations for $n>1$ were performed with the software package 
CADSOL/FIDISOL, based on the Newton-Raphson method \cite{FIDISOL}.
We use the general relations (\ref{total-energy})-(\ref{m-charge}) 
to compute the energy, angular momentum
and charges of these configurations.
The solutions with $n=1$ corresponds 
to spherically symmetric dyons with $H_1=H_3=\Phi_2=0,~H_2=H_4=w(r),
~\Phi_1=H(r),~H_5=J(r)$, 
and generalize the Julia-Zee solution \cite {Julia-Zee} for a negative cosmological constant.
Typical spherically symmetric solutions are presented in Fig. 1.

Although the results we find are broadly similar to those valid for the 
$\Lambda=0$ case, there are some differences.
The most interesting difference, not discussed in \cite{Lugo:1999fm}, 
is that the magnitude of the electric potential $A_t$
at infinity is no longer restricted.
In an asymptotically Minkowski spacetime, the constant $\alpha$ in eq.~(\ref{cond-inf}) is restricted to 
$\alpha \le 1$. 
For $\alpha>1$ some gauge field functions become oscillating instead of
asymptotically decaying. 
The $A_t^a$ components of the gauge filed act like an isotriplet Higgs field
with negative metric, and by themselves would cause the other components of the gauge field
to oscillate rather than decrease exponentially as $r \to \infty$ \cite{Julia-Zee,Hartmann:2000ja}.
However, in a AdS spacetime, finite energy solutions with arbitrary $\alpha$ are allowed.
The Higgs field forces $H_1=H_2=H_4=H_6=0$ asymptotically, but does not restrict the value of $H_5$.
In the limit of vanishing Higgs field ($\eta \to 0$)
we recover the (pure-) YM 
solutions discussed in \cite{Radu:2001ij, Hosotani:2001iz} (in the absence of the Higgs field
the value at infinity of the magnetic potentials $H_2,H_4$ is arbitrary).
Another distinctive feature of the AdS solutions is a smaller dyon radius 
as compared to the $\Lambda=0$ case. 

Apart from these properties, 
qualitatively, the Higgs field and Yang-Mills field behavior is very 
similar to that corresponding
to Minkowski spacetime dyons.
We notice a similar shape for the functions $H_i$ and $\Phi_i$ and also for the 
energy density. 
Increasing in size at the origin for increasing $|\Lambda|$, the energy density is localized in 
a decreasing region of space.
The energy of the multidyons is on order of $n$ times the corresponding one-dyon energy.
In Fig. 2 the energy and the electric charge of the solutions 
is plotted as a function of the cosmological constant 
for various winding numbers.
The toroidal shape familiar from the energy density of the dyon solutions in flat space 
is retained for solutions in AdS space, as is illustrated in Fig. 3.

Most of our results have been obtained in the Prasad-Sommerfeld limit, 
but similar conclusions hold for a nonvanishing Higgs self-coupling.
Further details on these solutions will be presented elsewhere.

Returning to the question of angular momentum,
by using the ansatz (\ref{ansatz-AdS}) with the boundary conditions (\ref{cond-inf}) we find again
\begin{equation} 
\label{res-dyon1} 
J=-2 \pi \lim_{r \rightarrow \infty} 
\int_{0}^{\pi} d \theta \sin \theta 
r^2 W^{(r)}F_{r t}^{(r)}=2\pi n Q_e\int_{0}^{\pi} d\theta \sin \theta \cos \theta=0.
\end{equation} 
It is worthwhile to notice that these solutions carry a nonvanishing angular momentum density, 
even in the Prasad-Sommerfeld limit, 
as ilustrated in Fig. 3c (see also Fig. 4).
The fact that $J=0$ can be  atributed again to the extra symmetries of the reduced ansatz, since 
as $\theta \to \pi-\theta$ we have $H_1\to -H_1,~H_2\to H_2,
~H_3\to -H_3,~H_4\to H_4,~H_5\to H_5,~H_6\to -H_6$
which implies the vanishing of the
integral (\ref{J}).
We find also that the YM dyon solutions discussed in \cite{Radu:2001ij} 
have a vanishing angular momentum.
 
Evidently, dyon solutions within the most general ansatz 
may also exist.
The construction of such solutions represents a
difficult challenge. 
However, we argue that even for the most general ansatz, an AdS  finite 
energy dyon solution presents a vanishing total angular momentum.

The asymptotic behavior of the functions $A_{\mu}^a,\Phi^a$ is complicated by the presence 
of a negative cosmological constant. 
Similar to the $n=1$ case \cite{Lugo:1999fm, Lugo:1999ai}, 
the transversal components of the YM potentials present
a complicated power decay, instead of the expected exponential law.  
Nevertheless, following the analysis in the $\Lambda=0$ case, we find that the 
required fall-off conditions for a
finite total energy are again  $F_{rt}^a \sim 1/r^{1.5+\epsilon}$, 
$W_{\theta}$ and $ W_{\varphi} \sim 1/r^{0.5+\epsilon}$,
while $A_{\varphi}^{r}$ shold also vanish for a regular configuration. 
Thus, after replacing in (\ref{T34}) we find again $J=0$.

\section{Further discussion}
We presented arguments that the results obtained within a perturbative approach in \cite{HSV}
remain valid 
in the general case, for the most general YMH axially symmetric ansatz. 
A flat space axially symmetric dyon configuration necessarily
possesses  a vanishing total angular momentum.

We emphasize the role played by the Higgs field boundary condition at infinity 
in obtaining
this nonexistence result.
We can of course consider a different behavior at infinity, consistent also with (\ref{Hinf})
and finite energy conditions, corresponding, however, to composed configurations. 
A sufficiently general asymptotic behavior 
of the Higgs field at infinity will read
\begin{equation}
\label{asa}
\Phi^r= \eta \cos f(\theta),~~~
\Phi^{\theta}= \eta \sin f(\theta),~~~\Phi^{\varphi}=0,
\end{equation} 
where $f(\theta)$ is an arbitrary function of $\theta$ reflecting a gauge freedom.
The regularity along the $z-$axis imposes that $\sin f(0)=\sin f(\pi)=0$.
Therefore we find two possible sets of boundary conditions: 
$\cos f(0)=\cos f(\pi)=\pm 1$
or
$\cos f(0)=-\cos f(\pi)=\pm 1$  (see also the discussion in \cite{Jang:hc}).
We call the first set even boundary conditions,
while the second set (for $f(\pi)=f(0)+(2k+1)\pi)$ will be refered to as
odd boundary conditions.

For the generic Higgs field boundary conditions (\ref{asa})
and within the reduced ansatz (\ref{ansatz-AdS}),
we find the asymptotic behavior in a flat space background
\begin{eqnarray}
\label{as1}
H_1&=&0,~~H_2=-f'(\theta),~~
H_5=\alpha \cos f(\theta),~H_6=\alpha \sin f(\theta),
\nonumber
\\
F_{rt}^{(r)}&=&\frac{ Q_e\cos f(\theta)}{r^2},~~F_{rt}^{(\theta)}=\frac{Q_e\sin f(\theta) }{r^2}.
\end{eqnarray} 
For even boundary conditions, the regularity 
and finite energy conditions impose the
asymptotic expression of the gauge potentials $H_3,~H_4$
\begin{eqnarray}
\label{as-even}
H_3=\frac{\cos \theta \cos f(0)}{\sin \theta}(\cos f(\theta) -\cos f(0)),~~
H_4=-\frac{\cos \theta}{\sin \theta}\sin f(\theta)\cos f(0), 
\end{eqnarray}
while for the second type of boundary conditions we find
\begin{eqnarray}
\label{as-odd}
H_3=\frac{\cos f(0)}{\sin \theta}(\cos f(\theta) -\cos \theta \cos f(0)),~~
H_4=-\frac{\cos f(0) \sin f(\theta)}{\sin \theta}. 
\end{eqnarray} 
By using these expressions and the relations (\ref{m-charge}), (\ref{totalJ}) 
we find that the even boundary configurations
have a magnetic charge $Q_m=\pm n$ and a vanishing total angular momentum.
The configurations satisfying odd-boundary conditions present a complementary picture: 
they posses a zero net magnetic charge and a nonzero angular momentum proportional to the 
electric charge
$J=\pm 4 \pi n Q_e$.
Thus, as conjectured in \cite{VanderBij:2001nm}, it seems that the vanishing of 
the total angular momentum 
is related to the presence of a net magnetic charge. 

At this stage, we remark that the ansatz (\ref{ansatz-AdS}) is invariant 
under a gauge transformation $U=\exp\{i\Gamma(r,\theta) \tau_{\varphi}^n/2\}$ \cite{Hartmann:2001ic}, 
and the Higgs field transforms as
\begin{eqnarray}
\label{trans}
\nonumber
\Phi^r &\to& ~~\cos \Gamma \Phi^r+\sin \Gamma \Phi^{\theta},
\\
\Phi^{\theta} &\to&  -\sin \Gamma \Phi^r+\cos \Gamma \Phi^{\theta},
\end{eqnarray} 
which implies $f \to f-\Gamma(\theta)$ in (\ref{asa}).
We may use this relation to set $f(\theta)=k \theta$  
in the Higgs field boundary conditions (with $k$ an integer number).

A number of well-known configurations  (in a flat space background) 
are obtained for this choice of $f$.
The dyon configurations discussed above correspond to $k=0$.
For $k=1$, we have as $r \to \infty$
\begin{eqnarray}
\label{asypt-MA1}
\Phi^r= \eta \cos \theta,~~~
\Phi^{\theta}= \eta \sin\theta,~~~\Phi^{\varphi}=0,
\end{eqnarray}
which implies the asymptotic behavior, within the reduced ansatz (\ref{ansatz-AdS})
\begin{eqnarray}
\label{asypt-MA2}
H_1&=&H_3=0,~~H_2=H_4=-1,
~~H_5=\alpha \cos \theta,~H_6=\alpha \sin \theta,
\nonumber\\
W^{(r)}&=& -n \cos \theta,~~~
W^{(\theta)}= -n \sin\theta,~~~W^{(\varphi)}=0,
\\
F_{rt}^{(r)}&=&\frac{ Q_e\cos \theta}{r^2},~~~F_{rt}^{(\theta)}=\frac{Q_e\sin \theta }{r^2}.
\nonumber
\end{eqnarray}
This corresponds to a monopole-antimonopole configuration, with a vanishing
net magnetic charge but a nonzero electric charge. 
This regular solution is discussed in \cite{map, Kleihaus:2000sx} for $n=1,~H_5=H_6=0$ 
and a different basis $\tau_a^n$ in spherical coordinates.
The authors of these papers use a 
different combination of $\tau_{\rho}^n $ and $\tau_z$, 
more suitable for these boundary conditions
($ \tau_{r}^n=\sin 2\theta~\tau_{\rho}^n+\cos 2\theta~\tau_z,~~
\tau_{\theta}^n=\cos 2\theta~\tau_{\rho}^n-\sin 2\theta~\tau_z
$).
In \cite{VanderBij:2001nm} it was found that the total angular momentum of this solution
is proportional to the electric charge.

For $k=2$ we find a configuration consisting in two monopoles 
and one antimonopole (we consider only the case $n=1$). 
This configuration is sustained by the boundary conditions at infinity
\begin{eqnarray}
\label{MAM1}
~\Phi^r&=& \eta \cos 2\theta,~~~\Phi^{\theta}=\eta \sin 2\theta,~~~\Phi^{\varphi}=0,
\end{eqnarray}
which imply asymptotically
\begin{eqnarray}
\label{MAM2}
W^{(r)}&=& - \cos \theta \cos 2\theta,~~~
W^{(\theta)}= -\cos\theta\sin 2\theta,~~~W^{(\varphi)}=0,
\nonumber\\
F_{rt}^{(r)}&=&\frac{Q_e\cos 2\theta }{r^2},~~~F_{rt}^{(\theta)}=\frac{Q_e\sin 2\theta }{r^2},
\end{eqnarray}
(corresponding to $H_1=0,~H_2=-2,~H_3=-\sin 2\theta,~H_4=-\cos 2 \theta-1,
~H_5=\alpha \cos 2\theta,~H_6=\alpha \sin 2\theta$ for the ansatz (\ref{ansatz-AdS})).
One can verify that this configuration has a unit  magnetic charge and a 
zero total angular momentum.

One may speculate on the existence of solutions for an arbitrary integer $k$, 
representing composed configurations, 
containing both monopoles and anti-monopoles.
It appears that further analysis of such problems will require at least
some use of numerical techniques.
The role played by the Higgs field topology 
and the extra-symmetries of the ansatz needs further study also.

The inclusion of a negative cosmological constant in a YMH theory
leads to the same general picture.
Numerical results have been presented
showing the existence of dyon solutions in fixed AdS spacetime.
However, similar to the flat space counterparts, these dyon solutions 
present also a vanishing total angular momentum.
We can consider of course more general configurations containing both monopoles and 
antimonopoles, by using the set of boundary conditions at infinity derived above.
In this case, it is possible to write a similar relation between the electric charge and the angular momentum.

We expect that the inclusion of gravity in this analysis will not change these conclusions,
since the gravitational field does not affect the behavior at infinity  of the YMH system.

Although the obtained results are in agreement with the perturbative approach,
we still are confronted by the problem of the physical mechanism preventing a 
configuration with net magnetic charge from rotating.
A more general proof, beyond a specific ansatz (possibly based on  topological arguments),
is clearly necessary.
\\
\\
{\bf Acknowledgement}
\\
One of the authors (E.R.) acknowledges an important discussion with Erick W\"ohnert.

This work was performed in the context of the
Graduiertenkolleg of the Deutsche Forschungsgemeinschaft (DFG):
Nichtlineare Differentialgleichungen: Modellierung,Theorie, Numerik, Visualisierung.

\newpage
{\bf Figure Captions}
\\
\\
Figure 1:
\\
Typical spherically symmetric dyon solutions for the same value of Higgs self-coupling constant 
$\lambda=1$ and different values of $\Lambda,~\alpha$.
The total energy $E$ (in units $4 \pi \eta$)  and the electric charge $Q_e$ are also  marked.
\\
\\
Figure 2:
\\
The energy per topological charge  $E/n$ in units $4 \pi \eta$ (Figure 2a) and the electric charge 
per topological charge $Q_e/n$ (Figure 2b) 
are plotted as a function of the cosmological constant
for dyon solutions  with $\alpha=0.2$ and $\lambda=0.5$.
\\
\\
Figure 3:
\\
The energy density $-T_{t}^{t}$ (Figure 3a), 
surfaces of constant energy density (Figure 3b), and 
$T_{\varphi}^t$ component of the energy-momentum tensor 
(in units of $4 \pi \eta $) (Figure 3c)
for a dyon solution with topological charge $n=2$, total energy 
$E=5.522$, electric charge
$Q_e=0.086$ and a vanishing Higgs self-coupling,
as a function of the dimensionless
compactified coordinates $\rho = \bar r \sin \theta$ and
$z = \bar r \cos \theta$. Here $\bar x = r/(1+r)$ is the coordinate 
used in the numerical calculations.
In Figure 3b, the inner surface correspond 
to a value of the energy density $\epsilon=4$, 
while the value for the outer surface is 9.5.
The cosmological constant is $\Lambda=-6$.
\\
\\
Figure 4:
\\
The $T_{\varphi}^t$ component of the energy-momentum tensor 
(in units $4 \pi \eta$) at $\theta=\pi/4$
for dyon solutions with $\alpha=0.2,~\lambda=0.5$, 
is shown for several values of the cosmological constant. The topological charge
of these solutions is $n=3$.

\newpage

\setlength{\unitlength}{1cm}

\begin{picture}(16,16)
\centering
\put(-2,0){\epsfig{file=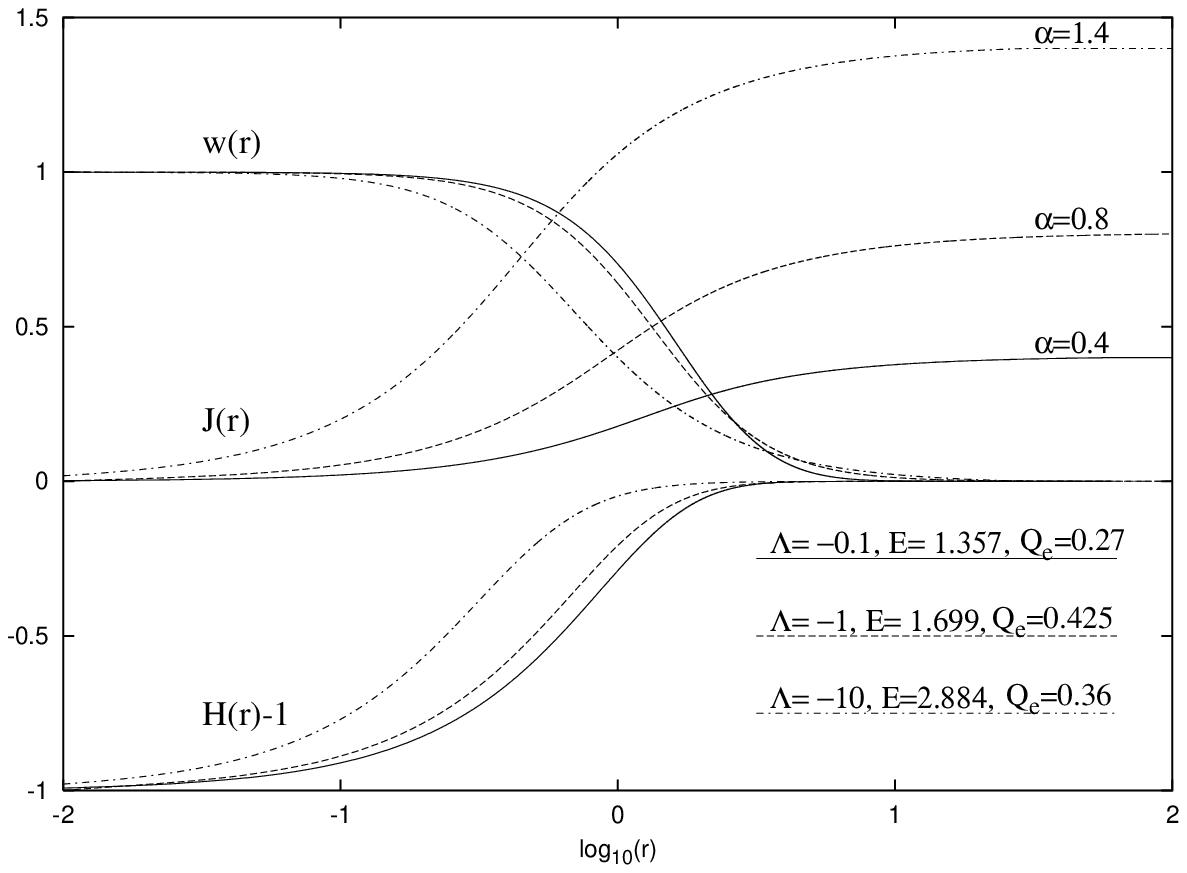,width=16cm}}
\end{picture}
\begin{center}
Figure 1.
\end{center}
\newpage

\setlength{\unitlength}{1cm}

\begin{picture}(16,16)
\centering
\put(-2,0){\epsfig{file=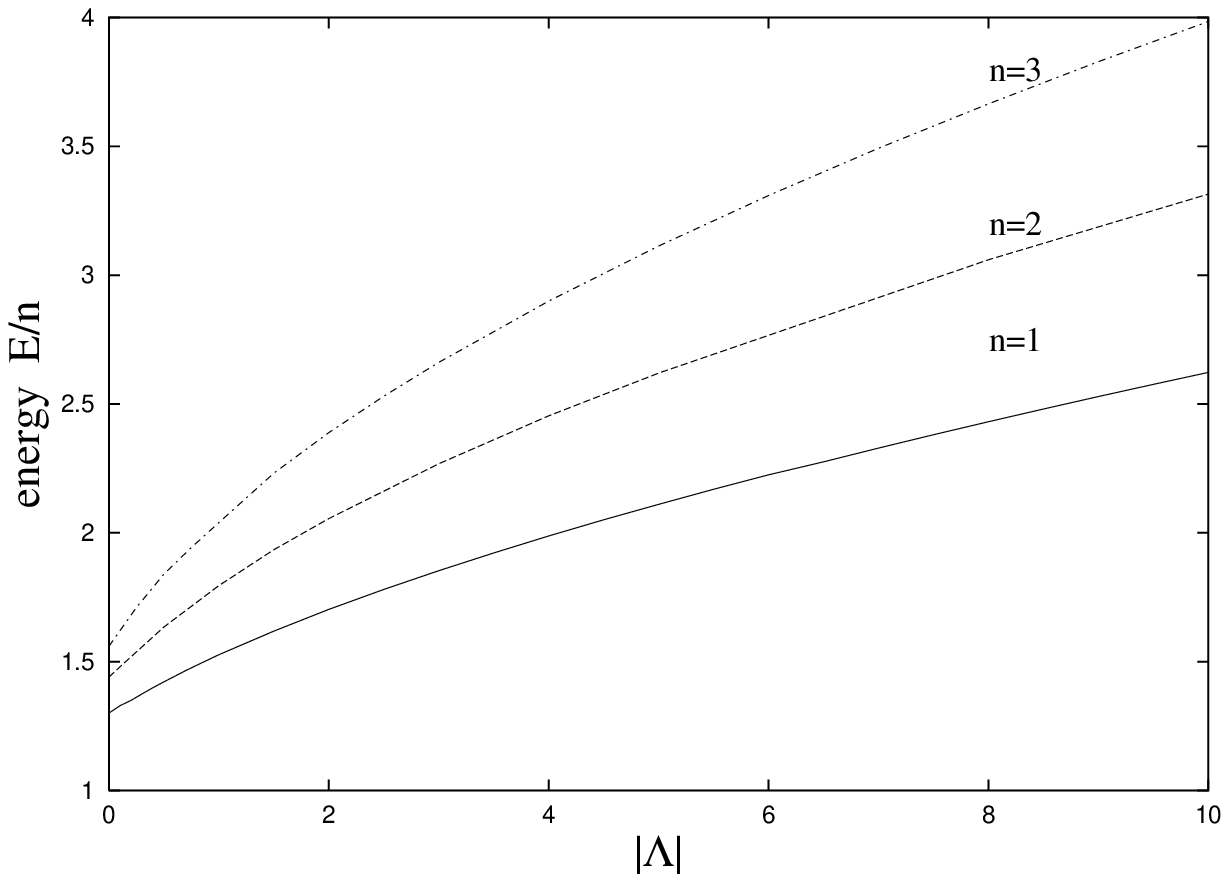,width=16cm}}
\end{picture}
\begin{center}
Figure 2a.
\end{center}

\newpage

\setlength{\unitlength}{1cm}

\begin{picture}(16,16)
\centering
\put(-2,0){\epsfig{file=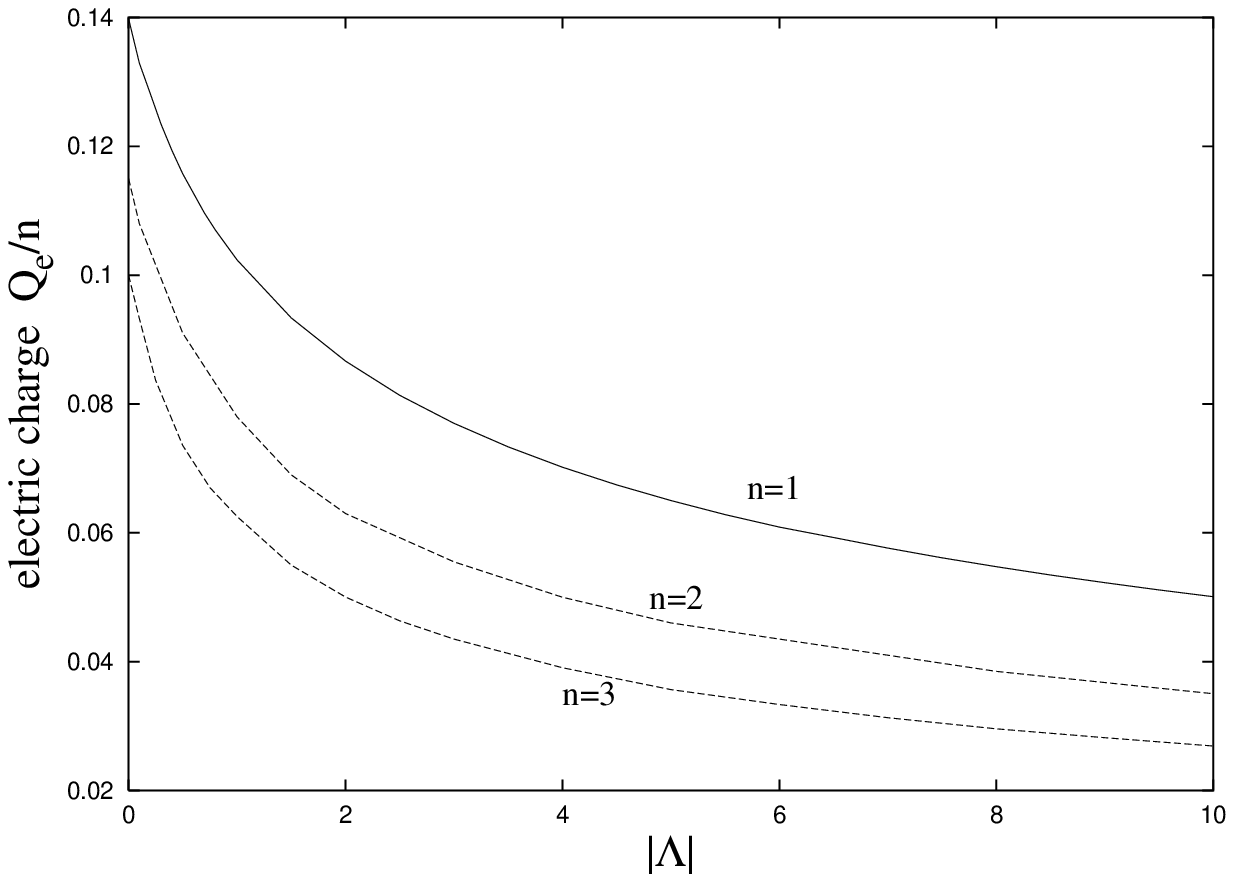,width=16cm}}
\end{picture}
\begin{center}
Figure 2b.
\end{center}

\newpage

\setlength{\unitlength}{1cm}

\begin{picture}(16,16)
\centering
\put(-2,0){\epsfig{file=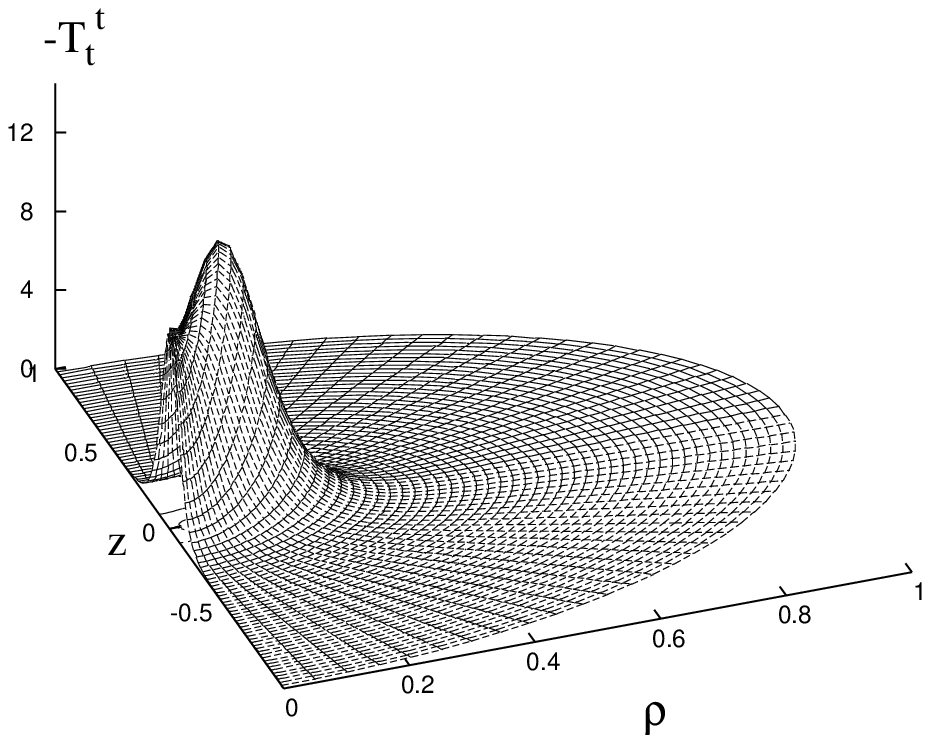,width=16cm}}
\end{picture}
\begin{center}
Figure 3a.
\end{center}

\newpage

\setlength{\unitlength}{1cm}

\begin{picture}(14,14)
\centering
\put(1,0){\epsfig{file=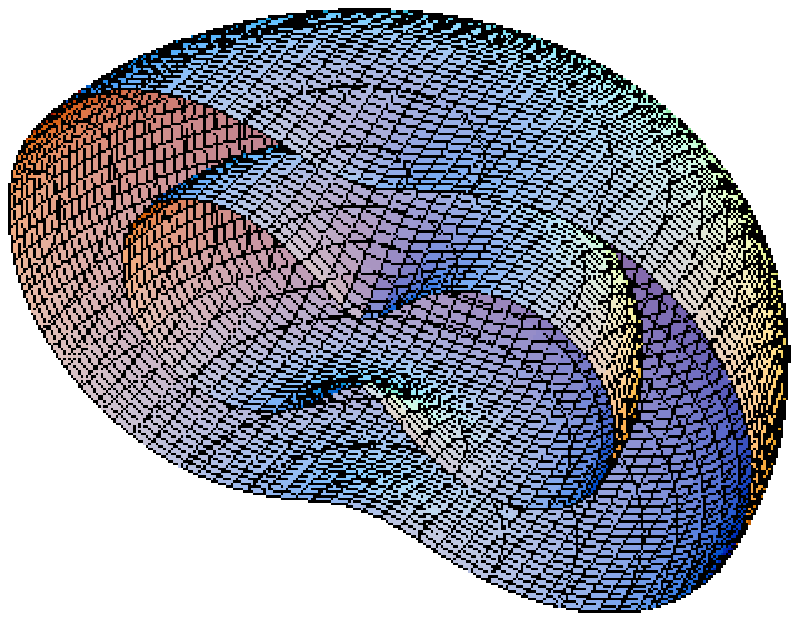,width=12cm}}
\end{picture}
\begin{center}
\vspace{3.cm}
Figure 3b.
\end{center}

\newpage

\setlength{\unitlength}{1cm}

\begin{picture}(16,16)
\centering
\put(0,0){\epsfig{file=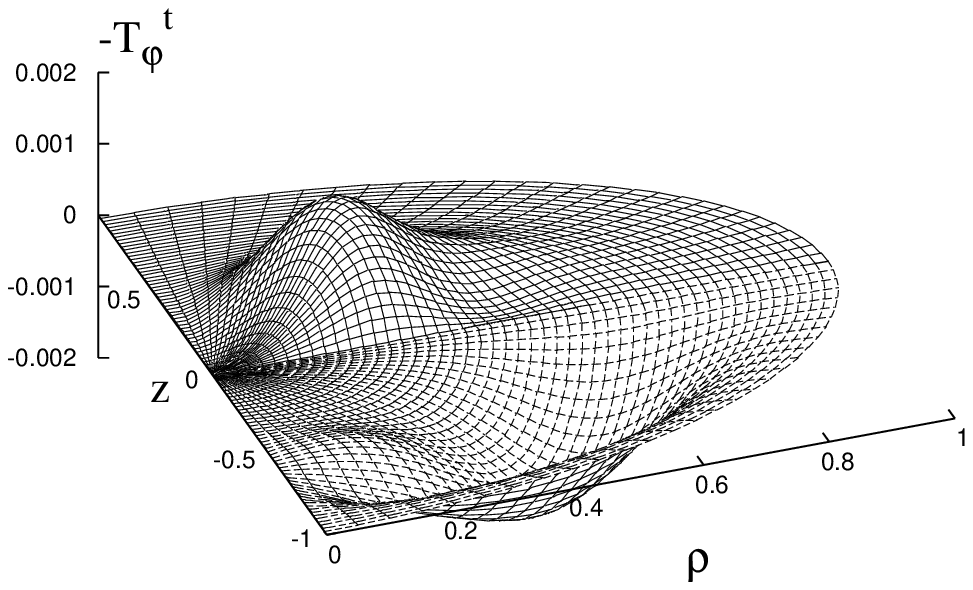,width=16cm}}
\end{picture}
\begin{center}
Figure 3c.
\end{center}

\newpage

\setlength{\unitlength}{1cm}

\begin{picture}(16,16)
\centering
\put(-2,0){\epsfig{file=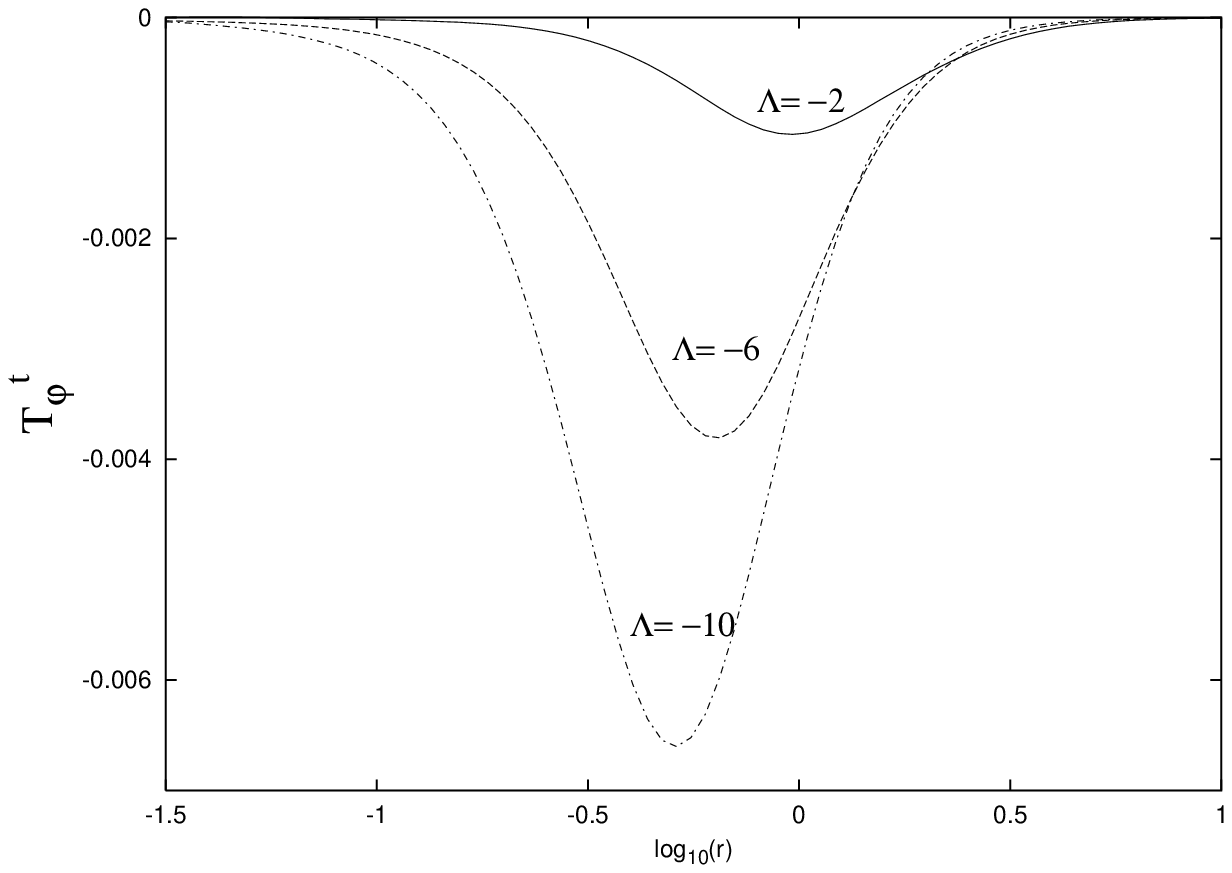,width=16cm}}
\end{picture}
\begin{center}
Figure 4.
\end{center}

\end{document}